\documentclass[]{aa}
\usepackage{txfonts}
\usepackage{graphicx}
\usepackage{natbib}

\begin{document}
\title{Inverse Compton X-rays from relativistic flare electrons and positrons}
\author{Alexander L. MacKinnon\inst{}\and Procheta C.V. Mallik \inst{}}

\institute {Department of Physics and Astronomy, University of
Glasgow, Glasgow G12 8QQ, U.K.}

\offprints{A.L. MacKinnon, \email{alec@astro.gla.ac.uk}}

\authorrunning{MacKinnon and Mallik}
\titlerunning{Inverse Compton flare X-rays}
\date{\today}

\abstract
{In solar flares, inverse Compton scattering (ICS) of photospheric photons might give rise to detectable hard X-ray photon fluxes from the corona
where ambient densities are too low for significant bremsstrahlung or recombination.
$\gamma$-ray lines and continuum in some large flares imply the presence of the necessary $\sim$ 100 MeV electrons and positrons, the latter as by-products of GeV energy ions.
Recent observations of coronal hard X-ray sources in particular prompt us to reconsider here the possible contribution of ICS.}
{We aim to evaluate the ICS X-ray fluxes to be expected from prescribed populations of relativistic electrons and positrons in the solar corona.
The ultimate aim is to determine if ICS coronal X-ray sources might offer a new diagnostic window on relativistic electrons and ions in flares.
}
{We use the complete formalism of ICS to calculate X-ray fluxes from possible populations of flare primary electrons and secondary positrons,
paying attention to the incident photon
angular distribution near the solar surface and thus improving on the assumption of isotropy made in previous solar discussions.}
{Both primary electrons and secondary positrons produce very hard ICS X-ray spectra. The anisotropic primary radiation field results in pronounced centre-to-limb variation in predicted fluxes and spectra, with the most intense spectra, extending to the highest photon energies, expected from limb flares. Acceptable numbers of electrons or positrons could account for RHESSI coronal X/$\gamma$-ray sources}
{Some coronal X-ray sources at least might be interpreted in terms of ICS by relativistic electrons or positrons, particularly when sources
appear at such low ambient densities that bremsstrahlung appears implausible.}


\maketitle

\section{Introduction} \label{sect:intro}

\cite{1967SvA....11..258K, 1971SoPh...18..284K} considered three possible radiation mechanisms via which solar
flare energetic electrons might produce hard X-rays (HXRs): synchrotron, bremsstrahlung and inverse Compton scattering
(ICS). He established that fluxes from electron-ion bremsstrahlung would dominate those from the other two mechanisms
under normal solar atmosphere conditions and thus laid one of the foundations of the interpretation of flare X-rays.
Left open, however, was the possibility that ICS HXR fluxes from low-density regions might exceed those from bremsstrahlung (or, indeed,
recombination - \cite{2008A&A...481..507B, Brown&Mallik:2009}). Recent
years have seen increasingly detailed observations of coronal HXR sources \citep{2001ApJ...561L.211H, 2008A&ARv..16..155K, 2008ApJ...678L..63K,  2009A&A...502..665T}, sometimes from surprisingly tenuous regions. Reconsideration of the possible role
of ICS in HXR production thus seems timely \citep{2008A&ARv..16..155K}.

The basics of ICS are well understood \citep[e.g.][]{1970RvMP...42..237B, 1986rpa..book.....R}. Suppose that electrons of (total) energy
$\gamma m_e c^2$ scatter photons of initial energy $\epsilon_i$. Optical photons of photospheric origin, for instance,
would have $\epsilon_i$ typically of order 2 eV. The maximum possible scattered photon energy results from a head-on
collision of electron and photon and has a value of $\epsilon_{max} \simeq 4\gamma^2 \epsilon_i$ \citep[e.g.][]{1970RvMP...42..237B}. To
produce HXR photons via ICS of optical photons thus needs electrons in the 10s to 100s of MeV energy range.

There is good evidence that electrons attain such energies in flares. $\gamma$-ray continuum
in this energy range has been observed from some large flares \citep[e.g.][]{1983Natur.305..291F, 1993A&AS...97..349K, 1993SoPh..147..137T}.
This may be due to either or both of: electron-ion bremsstrahlung from primary accelerated electrons; bremsstrahlung from secondary
electrons and positrons in the 100 MeV energy range, produced in reactions of accelerated ions in the energy range $>$ 0.3 GeV \citep[e.g.][]{1987ApJS...63..721M}.
In the latter case, positrons are dominant in number since they result from collisions between positively charged particles. Continuum
in this case is unavoidably accompanied by the flat spectral feature around 70 MeV produced by $\pi^0$ decay. High-energy continuum
can occur both with and without this feature at different times during a single event \citep[e.g.][]{2003A&A...412..865V}, indicating that
both primary accelerated electrons and secondary positrons may be present in the 100 MeV energy range, as needed for ICS HXR production.
\cite{1994AIPC..294..130A} give evidence that the flare of 26 March 1991 accelerated electrons to energies of 300 MeV.
The energy distributions of electrons and positrons will be very different, however, and we consider them separately.

In the presence of the solar magnetic field, these high energy electrons would also produce synchrotron emission but at radio and sub-mm wavelengths
\citep{2007SoPh..245..311S}. To produce X-rays by synchrotron emission would require electrons of unrealistically high energy, for which
there is no evidence.

The ICS estimates of \cite{1967SvA....11..258K, 1971SoPh...18..284K} and \cite{2008A&ARv..16..155K} employ standard results based on assuming isotropic electron
and photon distributions. Electron distributions in the corona may well be isotropic because of pitch-angle scattering by MHD
turbulence \citep[e.g.][]{1989ApJ...344..973M} but the photon distribution will be isotropic only in the outward hemisphere. As already
mentioned, the most energetic photons result from head-on collisions of photon and electron, which result in the up-scattered photon
travelling along the direction of the incident electron \citep{1968PhRv..167.1159J}. These most favourable collisions clearly cannot occur,
even with an assumed isotropic coronal electron distribution, so a more involved calculation is essential to evaluate likely
ICS fluxes, spectra etc.

ICS is certainly important in other areas of astrophysics: of cosmic microwave background photons by hot gas in clusters of
galaxies \citep[Sunyaev-Zeldovich effect -][]{1970Ap&SS...7....3S}; of solar visible photons by cosmic ray electrons \citep{2008A&A...480..847O}.

The formalism for calculating ICS radiation with arbitrary photon angular distributions has been given most recently
by \cite{2000ApJ...528..357M}. Here we adapt their work to the source geometry near the solar surface. We use typically
observed power-law distributions of electrons and protons (which produce secondary positrons) to illustrate our study. We elucidate the
consequences for observability of this ICS flux and note the difference between the spectra produced by electrons and by secondary positrons,
as well as the disc-centre to limb variation. Our findings reveal that although the ICS intensities are likely to be low, the spectrum is hard
and unmistakable. If detected by modern instruments, this would be a new window on extremes of electron and ion acceleration at the Sun, and in the case of ions complementing information available from $\gamma$-ray lines and free neutrons detected in space.

In this paper, we use the units $\hbar = c = m_e = 1$.

\section{Source geometry; calculation of IC flux}

In this section we calculate ICS HXR fluxes from relativistic electron and positron populations in the corona, following \cite{2000ApJ...528..357M}.

The rate of photon-particle interactions is given in full generality by \citep{1976PhRvA..13.1563W}:

\begin{equation}
R = n_en_{\gamma}\int d\overrightarrow{p}_{\gamma} \int d\overrightarrow{p}_e f_e(\overrightarrow{p}_e) f_{\gamma}(\overrightarrow{p}_{\gamma}) \frac {p_{\gamma}'}{\gamma p_{\gamma}} \sigma (p_{\gamma}'),
\label{weaver}
\end{equation}
\noindent
where $n_e$, $n_\gamma$ are the electron and photon number densities; $\overrightarrow{p}_e$, $\overrightarrow{p}_{\gamma}$ are the momenta;
$f_e(\overrightarrow{p}_e)$,  $f_{\gamma}(\overrightarrow{p}_{\gamma})$ are the respective distribution functions in the laboratory system (LS),
normalised to unity; $\gamma$ is the electron Lorentz factor; $\sigma$ the cross-section; and the primes signify the electron rest system (ERS) variables.
For relativistic electrons, the incoming photons are seen as a narrow beam $\sim 1/\gamma$ wide in the ERS. We follow \cite{2000ApJ...528..357M} in using
\cite{1968PhRv..167.1159J} approximation that the incident photons are seen as a unidirectional beam in the ERS. This significantly simplifies
the calculation of the ICS fluxes while introducing negligible error \citep{1968PhRv..167.1159J}.

To calculate fluxes from Equation~(\ref{weaver}) we need to specify the electron and photon momentum distributions and the cross-section. Since we
deal with highly relativistic particles and situations where the photon may carry away a large fraction of the electron energy, we must use the
Klein-Nishina cross-section e.g. as given by \citep{1976tper.book.....J}:
\begin{eqnarray}
\nonumber
\frac {d\sigma}{d\epsilon_2'\, d \cos\eta'} = \pi r_e^2 \left(\frac {\epsilon_2'}{\epsilon_1'}\right)^2 \left(\frac {\epsilon_2'}{\epsilon_1'} + \frac {\epsilon_1'}{\epsilon_2'} - \sin^2\eta' \right) \\
\times \, \delta \left[\epsilon_2' - \frac {\epsilon_1'}{1 + \epsilon_1'(1 - \cos\eta')}\right],
\label{knxs}
\end{eqnarray}
\noindent
where $r_e$ is the classical electron radius, $\epsilon_1'$ and $\epsilon_2'$ are the ERS energies of the incident and up-scattered photons, $\eta'$ is the scattering angle in the ERS and $\delta (x)$ denotes the Dirac delta function.

Appropriately, to the general galactic cosmic ray population, \cite{2000ApJ...528..357M} assume isotropic electrons;
this assumption will also be appropriate in the corona as a result of MHD scattering \citep{1989ApJ...344..973M, 1992ApJ...389..739M}. With these assumptions,
the up-scattered photon distribution over the LS energy, $\epsilon_2$, as obtained from Equation~(\ref{weaver}) is \citep{2000ApJ...528..357M}

\begin{eqnarray}
\frac{dR}{d\epsilon_2} = \int d\cos\eta' \int d\epsilon_1 d\Omega_\gamma \int d\gamma d\Omega_e \\
\nonumber
\times f_e(\gamma,\Omega_e) f_\gamma(\epsilon_1, \Omega_\gamma) \epsilon_1^2 \gamma^2 \frac {\epsilon_1'}{\gamma \epsilon_1} \frac {\epsilon_2'}{\epsilon_2} \frac {d\sigma}{d\epsilon_2'd\cos \eta'},
\label{msflux}
\end{eqnarray}
\noindent
where $\Omega_\gamma$ and $\Omega_e$ refer to photon and electron directions respectively.

At this point we depart from \cite{2000ApJ...528..357M}, tailoring our calculation to the radiation field geometry above the
solar surface (Figure 1). We introduce two, spherical polar angular coordinates $\theta$ and $\phi$ to label photon direction. Let $\hat{n}$
be a unit vector pointing radially outward from the local solar surface, and $\hat{l}$ be a unit vector pointing along the line of sight to the
observer. Then we have $\hat{l}.\hat{n} = \sin\lambda$ where $\lambda$ is the heliocentric angle of the source location. Let $\hat{p}_\gamma$ be
a unit vector in the direction of the
photon. The polar angle $\theta$ measures the angle between $\hat{n}$ and $\hat{p}_\gamma$, i.e. $\hat{n}.\hat{p}_\gamma = \cos\theta$. The photon
azimuthal angle $\phi$ lies in the plane of the solar surface and is measured anticlockwise from the plane defined by $\hat{n}$ and $\hat{l}$.

\begin{figure}
\includegraphics[width=0.4\textwidth]{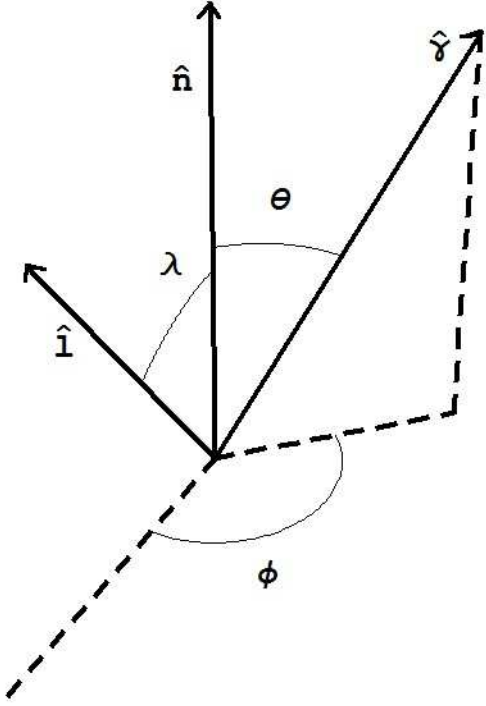}
\caption{Schematic diagram showing the geometry used to describe the radiation field at the solar surface with the relevant angles and vectors. $\phi$ lies in the
solar surface plane.}
\end{figure}

The photon distribution is isotropic in the optically thick photosphere but only includes outward-flowing photons immediately above. It will be close to
isotropic, in the hemisphere $\theta < \pi/2$, as long as we consider coronal locations below $\sim 2R_{\odot}$. Thus the photon angular
distribution takes the simple form
\begin{equation}
f_\gamma(\epsilon_1, \theta, \phi ) \, = \, \frac{1}{2 \pi} H\left(\frac{\pi}{2}-\theta\right) g_\gamma(\epsilon_1),
\label{photdist}
\end{equation}
where $H$ is the Heaviside step function.

In the first instance we calculate the ICS flux from monoenergetic electrons with a single energy $\gamma$, averaging straightforwardly over more general
energy distributions as needed. We also consider monoenergetic primary photon distributions, $g_\gamma(x)=\delta(x-\epsilon_1)$. Using Equations~(\ref{knxs}), (3) and (\ref{photdist}), we hence find the total up-scattered photon distribution,
per electron, over the LS energy, $\epsilon_2$, to be:

\begin{eqnarray}
\frac {dR}{d\epsilon_2} = \left(2 - \frac {4\epsilon_2}{\gamma} + \frac {3\epsilon_2^2}{\gamma^2} - \frac {\epsilon_2^3}{\gamma^3} \right) \int_0^{2\pi} \int_{\theta_{min}}^{\theta_{max}} \sin \theta \, d\theta \, d\phi \\
\nonumber
- \frac {1}{\epsilon_1\gamma} \left(\frac {2\epsilon_2^2}{\gamma^2} - \frac {2\epsilon_2}{\gamma} \right) \int_0^{2\pi} \int_{\theta_{min}}^{\theta_{max}} \frac {d\cos \theta}{1 + \cos \theta} d\phi \\
\nonumber
- \frac {\epsilon_2^2}{\epsilon_1^2 \gamma^4} \int_0^{2\pi} \int_{\theta_{min}}^{\theta_{max}} \frac {d\cos \theta}{(1 + \cos \theta)^2} d\phi.
\label{pmflux}
\end{eqnarray}
\noindent

The lower limit of the $\theta$ integral is given by kinematics:
\begin{equation}
\theta_{min} = \arccos \left(1 - \frac {\epsilon_2}{2\epsilon_1 \gamma (\gamma - \epsilon_2)}\right)
\end{equation}
\noindent
and the upper limit by source geometry:
\begin{equation}
\theta_{max} = \arccos (\sin \lambda \cos \phi).
\end{equation}
\noindent
Performing the integral over polar angle we get

\begin{eqnarray}
\noindent \nonumber
\frac{dR}{d\epsilon_2}=\frac{r_e^2}{2\epsilon_1(\gamma - \epsilon_2)^2} \times \\
\nonumber
\int_0^{2\pi}\left(\frac{\epsilon_2^3}{\gamma^3} - \frac{3\epsilon_2^2}{\gamma^2} + \frac{4\epsilon_2}{\gamma}-2\right)\left(\sin\lambda \cos\phi - \left(1-\frac{\epsilon_2}{2\epsilon_1\gamma(\gamma-\epsilon_2)}\right)\right)\\
\nonumber
+\frac{2\epsilon_2}{\epsilon_1\gamma^2}\left(1-\frac{\epsilon_2}{\gamma}\right)\left(\ln(1+\sin\lambda \cos\phi)-\ln\left(2-\frac{\epsilon_2}{2\epsilon_1\gamma(\gamma-\epsilon_2)}\right)\right)\\
+ \frac{\epsilon_2^2}{\epsilon_1^2\gamma^4}\left(\frac{1}{1+\sin\lambda \cos\phi}-\frac{1}{2-\epsilon_2/(2\epsilon_1\gamma(\gamma-\epsilon_2))}\right)d\phi,
\label{finflux}
\end{eqnarray}

\begin{figure}
\includegraphics[width=0.55\textwidth]{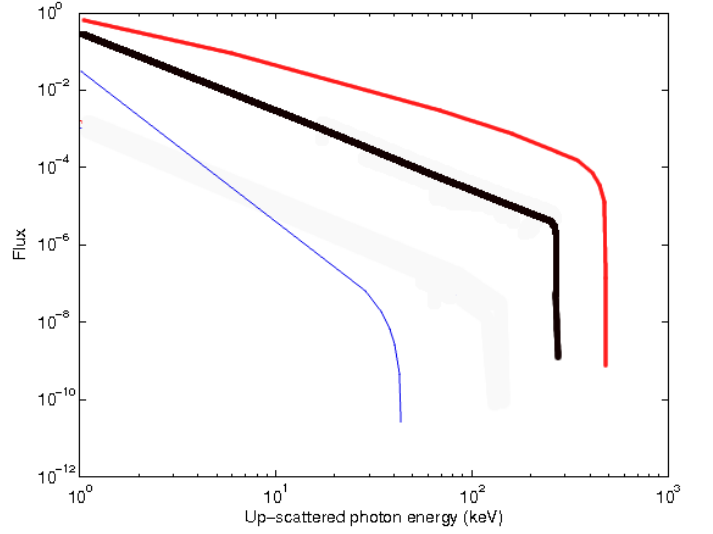}
\caption{Photon spectra at the Sun (photons per keV per second per source electron) from limb fast electrons with different power-law distributions: thin-black is for $\delta=5$, medium-red for $\delta=3$ and thick-blue for $\delta=2$ with an incident photon energy of 2 eV.}
\end{figure}

\begin{figure}
\includegraphics[width=0.55\textwidth]{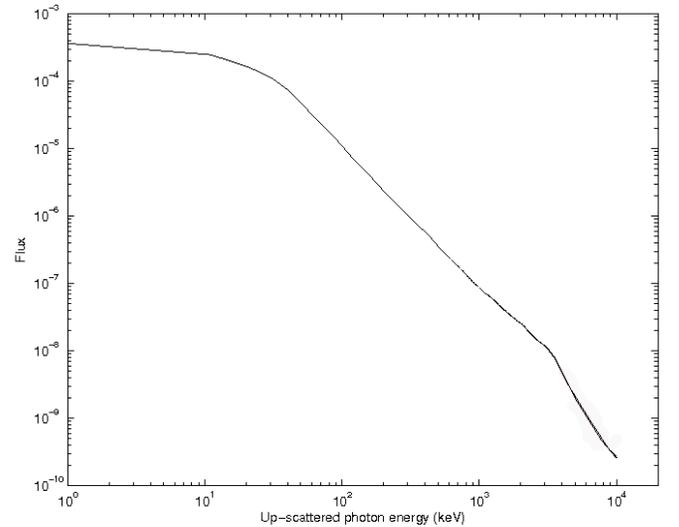}
\caption{Photon spectra at the Sun (photons per keV per second per source electron) from fast electrons with a power-law energy distribution $E^{-3}$ for an incident photon energy of 200 eV.}
\end{figure}

\noindent
which is the ICS flux of photons per unit energy per unit time per electron.
The following kinematic results \citep{2000ApJ...528..357M} are also of importance:

\begin{equation}
\epsilon_2' = \epsilon_2/[\gamma(1-\cos\eta')], ~~ \epsilon_2\le2\gamma\epsilon_1'/(1+2\epsilon_1'), ~~ \epsilon_1' = \epsilon_1\gamma(1+\cos\lambda).
\end{equation}

The maximum energy of the up-scattered photon is

\begin{equation}
\epsilon_{2~max} = 4\epsilon_1\gamma^2/(1+4\epsilon_1\gamma).
\end{equation}

Note that the second and third terms in Equation~(\ref{finflux}) have to be evaluated numerically. This was done using MATLAB and the results are
portrayed in the following section.

\section{Results}

\subsection{ICS from fast electrons}

To calculate ICS spectra produced by relativistic electrons, we assumed power-law primary electron kinetic energy distributions extending into the 10s of MeV range,
$\sim (\gamma-1)^{-\delta}$. The incident photon population was assumed to have a monoenergetic energy distribution at $\epsilon_1 = 2$ eV (or 200 eV in a few,
illustrative cases) so that the solar luminosity implies a photon density $n_\gamma = 10^{12}$ cm$^{-3}$. We have checked that the spectra found using the full,
black body photospheric spectrum are not significantly different from these shown here for the 2 eV case.

In Figure 2, we show the ICS spectra from electrons with energy spectral index $\delta = 3$,
calculated by weighting the emissivity (\ref{finflux}) by this distribution and integrating over electron energy. Fluxes are normalised
to one electron above 0.5 MeV and we assume an upper cutoff energy of 100 MeV.
The three separate curves signify the ICS spectrum as seen from three different viewing angles $\lambda$. In Figure 3, we show the ICS spectra for an event viewed at the limb but now for different values of $\delta$.

Clearly visible (2 eV) photons can easily be up-scattered to 10s of keV, even though the actual fluxes and spectra depart
from those expected on the basis of an isotropic photon distribution. Over most of the photon energy range the spectra are described by the expected
\citep[e.g.][]{1970RvMP...42..237B} ICS
power-law $\sim \epsilon_2^{-(1+\delta)/2}$, but falling off much more steeply as they approach an upper cutoff determined by the 100 MeV
electron upper cutoff, the viewing angle and the energy spectral index $\delta$. As expected on geometrical grounds, the most energetic
photons come from limb events. In the photon energy range produced across the disc, ICS exhibits pronounced limb-brightening with flux variations
of two orders of magnitude between identical events viewed at the limb and at disc centre. In Section~\ref{cfobs}
we see that observed coronal source photon fluxes imply plausible electron numbers.

Comparison of the ICS fluxes of Figures 2 and 3 with bremsstrahlung from the same electrons is not quite straightforward.
For the usual, monotonic declining energy
distributions of electrons, the bremsstrahlung flux at photon energy
$\epsilon$ is dominated by electrons with energies just above
$\epsilon$. ICS hard X-ray photons, however, are produced by electrons
in the 10s - 100s of MeV energy range. Any comparison of bremsstrahlung
and ICS fluxes involves an assumption about the electron energy
distribution over a very wide range. There is, for instance, evidence
that electron distributions routinely harden between 10s of keV and the
MeV energy range \citep[e.g.][]{2000ApJ...545.1116S}. Including a bremsstrahlung
spectrum for comparison in Figure 2 could be quite misleading in
consequence.

\begin{figure}
\includegraphics[width=0.55\textwidth]{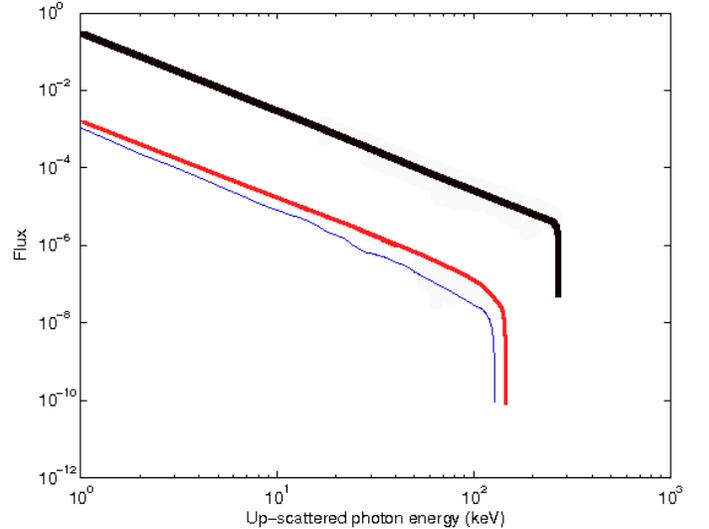}
\caption{Photon spectra at the Sun (photons per keV per second per source electron) from fast electrons with a power-law energy distribution $E^{-3}$,
where the thin-blue curve is the flux from disc centre ($\sin\lambda = 0$), medium-red for $\sin\lambda = 0.5$ and thick-black for the solar limb
($\sin\lambda = 1$). Fluxes
are normalised to one electron above 0.5 MeV and for an incident photon of energy 2 eV.}
\label{icelectron}
\end{figure}

For illustration, we may nonetheless assume that a single power law distribution $\sim (\gamma-1)^{-\delta}$ in
kinetic energy characterises the electron distribution all the way from
10 keV to 100s of MeV. Adopting $\delta = 3$, for example, we find that the bremsstrahlung flux at 10 keV
will be comparable to the ICS flux for an ambient density of about 10$^{10}$ cm$^{-3}$. The harder ICS flux will dominate at photon energies above this value,
until we approach the upper cutoff shown in Figure~\ref{icelectron}. Thus ICS appears likely to dominate
over bremsstrahlung for much of the time in the corona.

Still higher photon energies will result from primary photons of higher energy. For illustration we show in Figure 4 the spectrum resulting from ICS of primary
EUV photons of energy 200 eV, from a flare at disc centre. For easy comparison with the results for optical photons we have adopted the same photon density, $n_\gamma = 10^{12}$ cm$^{-3}$, although the true EUV density will be many orders of magnitude smaller - see below.

\subsection{ICS from relativistic positrons}

As noted in Section 1, positrons will be produced as secondaries from fast ion reactions. Electrons and positrons with the same energy
distribution would of course produce identical ICS spectra, but the positron energy distribution from p-p collisions, and hence
the ICS photon spectrum, is quite different from the power-law electron case considered in Section 3.1. We calculate positron energy distributions as in
\cite{2003A&A...412..865V}, which in turn closely follows \cite{1986ApJ...307...47D, 1986A&A...157..223D}, assuming they are produced via
pion decay following reactions of fast protons with ambient H and He nuclei. The nuclear reactions producing the positrons occur mostly
in the chromosphere and photosphere, but with a range of directions. At the energies considered here, any that mirror above the photosphere will
suffer only an insignificant energy loss as they make their way into the corona \citep{1990A&A...232..544M}. For simplicity, we assume here that
we may use the positron energy distribution from pion decay unmodified by any other processes. A more detailed treatment of transport will be carried
out elsewhere. We see in Section~\ref{cfobs} that only a few percent of the number of positrons produced in a large flare will give a detectable ICS source.

Positrons may also be produced in flares via beta decay of unstable nuclei produced in nuclear
reactions of flare ions. As noted by \cite{1987ApJ...316..801K}, positrons produced in this way generally have energies $< 1$ MeV, too low to be of interest here.

ICS spectra from the resulting positrons are shown in
Figure 5, assuming a power-law proton energy distribution with $\delta=3$ extending to an upper cutoff energy of 3 GeV and, again, $\epsilon_1$ = 2 eV and $n_{\gamma} = 10^{12}$ cm$^{-3}$.  Secondary positron distributions have a maximum at
about 300 MeV and a form that is dominated by the nuclear physics of pion formation and decay until primary proton energies significantly exceed
the threshold for pion production \citep{1987ApJS...63..721M}. Thus the detailed photon spectra depend rather weakly on proton power-law energy spectral index. However, certain features persist, i.e. the spectrum remains very hard and the most energetic photons will once again come from limb events. The three separate curves are for three different values of viewing angle $\lambda$. Also shown is the dashed-green curve in Figure 5, which is the bremsstrahlung spectrum from the same positrons, assuming an ambient density of $10^{10}$ cm$^{-3}$. We
used the cross-section of Bethe and Heitler, without making non-relativistic or extreme
relativistic approximations \citep[][ - formula 3BN]{1959RvMP...31..920K} and the relativistic electron-electron cross-section of \cite{1998SoPh..178..341H}, noting that electron-electron and electron-positron cross-sections become identical for relativistic energies \citep{1985PhRvD..31.2120H}. As mentioned above, the form of the positron distribution depends rather weakly on assumptions about the primary ion distribution so this comparison can be made with much more certainty than for electrons. Even with this ambient density, fairly high for the corona, ICS dominates over the bremsstrahlung flux from the same positrons. Annihilation of positrons in flight yields a continuum photon flux that may be neglected compared to bremsstrahlung, for present purposes \citep{1987ApJS...63..721M}. 

In Figure 6, we show the ICS spectra for a range of proton energy distribution $\delta$ values.
Secondary positron typical energies naturally result in up-scattering to the MeV photon energy range.

As for the electron case, we would expect a more energetic ICS flux if we consider incident EUV photons, shown in Figure 7 for 200 eV incident photons. With the photon density held fixed, as for Figure 5, the ICS flux can be as much as four orders of magnitude greater for $\epsilon_1 =$ 200 eV than
for $\epsilon_1 = $2 eV, at the same time extending to higher energies. So we would need an EUV photon density $\sim 10^{-4}$ times that of visible photons to produce an equally intense ICS flux. A rough estimate of EUV photon density in a large flare suggests this will be $\sim 10^3$ cm$^{-3}$, however, so low that even the greater fluxes obtained with more energetic incident photons will not be observable.

\begin{figure}
\includegraphics[width=0.55\textwidth]{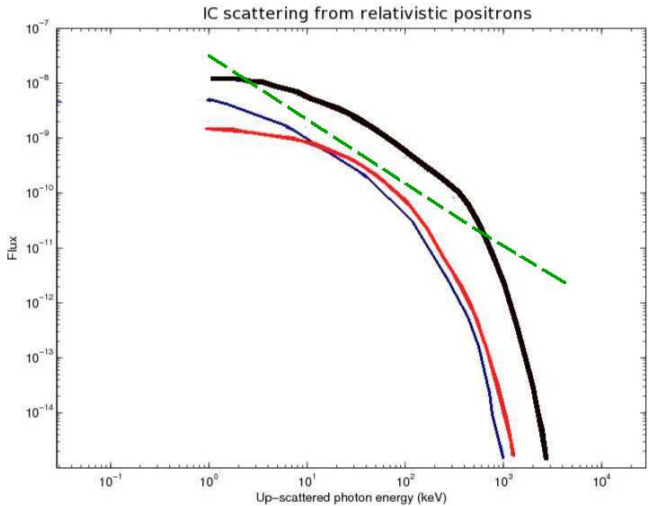}
\caption{Photon spectra at the Sun (photons per keV per second per proton) from relativistic positrons produced by protons with a power-law energy distribution $E^{-3}$, where the thin-blue curve is the flux from disc centre ($\sin\lambda = 0$), medium-red for $\sin\lambda = 0.5$ and thick-black for the solar limb flux ($\sin\lambda = 1$). Fluxes are normalised to one proton above 1 MeV and for an incident photon energy of 2 eV. The dashed-green line represents the bremsstrahlung flux from the same positron distribution and has been included here to compare with predicted ICS fluxes. Note that due to the extreme relativistic nature of the positrons, the bremsstrahlung flux is hard with an index of 1. In other words, for such energetic positrons, their distribution has little effect on the spectral index of the bremsstrahlung radiation they produce.}
\end{figure}

\begin{figure}
\includegraphics[width=0.55\textwidth]{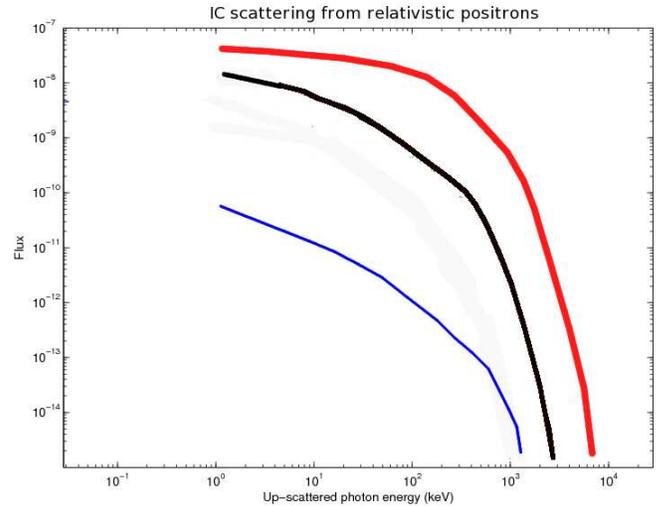}
\caption{Photon spectra at the Sun (photons per keV per second per proton) from limb relativistic positrons produced by protons with different power-law distributions: thin-blue is for $\delta=5$, medium-black for $\delta=3$ and thick-red for $\delta=2$ with an incident photon energy of 2 eV.}
\end{figure}

\begin{figure}
\includegraphics[width=0.55\textwidth]{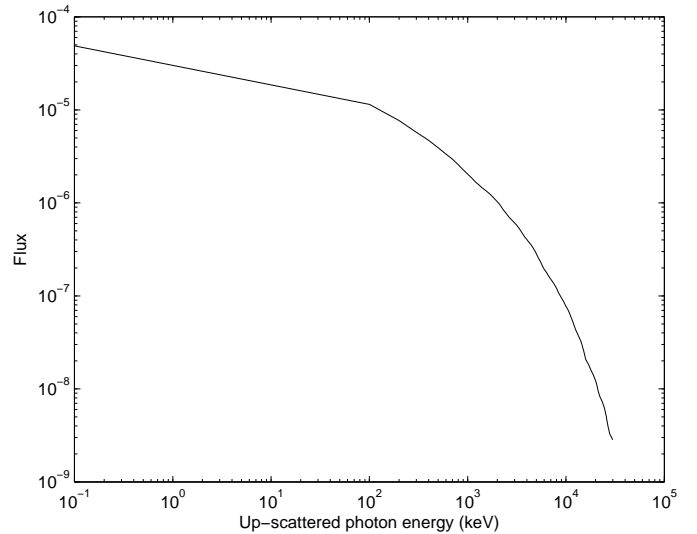}
\caption{Photon spectra at the Sun (photons per keV per second per proton) from relativistic positrons produced by protons with a power-law energy distribution $E^{-3}$ for an incident photon energy of 200 eV.}
\end{figure}

\section{Comparison with observations}
\label{cfobs}
As explained in Section 1, ICS could be dominant in producing HXRs in low-density regions of the solar atmosphere, which mainly implies the high corona.
Consider the coronal X/$\gamma$-ray source in the 2005 January 20 flare, described by \cite{2008ApJ...678L..63K}. Could it be due to ICS of photospheric photons?

Continuum $\gamma$-radiation in the 100 MeV energy range was observed from this flare by the SONG instrument on CORONAS-F. There is evidence for a pion decay
contribution to the observed spectrum \citep{2005ICRC....1...49K}, which would also
indicate the presence of $\sim 100$ MeV positrons. The flare was located towards the limb (N14$^\circ$W61$^\circ$; $\sin\lambda = 0.88$), maximising the likelihood of observable
ICS photons. Moreover, the location of the coronal X-ray source is high enough that ICS could be the dominant source of HXRs, given sufficient energetic electrons.
 The coronal source has a very hard spectrum, photon spectral index $\approx 1.5$, consistent with the
spectra found in Section 3. A photon spectral index of 1.5 would imply a relativistic electron spectral index of about 2. Continuation of the photon spectrum
to at least 700 - 800 keV implies an electron distribution continuing in this power-law form to at least 120 MeV.
To account for the observed coronal
source fluxes shown in Figure 3 of \cite{2008ApJ...678L..63K}, we would need $\sim 10^{31}$ electrons instantaneously present
above 0.5 MeV. The $\sim 500$ keV source represented by the 50\% contour of \cite{2008ApJ...678L..63K}, Figure 2c, is about 40$\times$80 arc seconds. Assuming
a similar length scale along the line of sight we estimate its total volume as $5 \times 10^{28}$ cm$^3$. Taking, for illustration, an
ambient electron density $10^8$ cm$^{-3}$, we see that the relativistic electrons necessary to account for this source via ICS would represent just
$2\times 10^{-6}$ of all particles in the volume. We also estimate that this is $\sim10^{-3}$ or less of the electrons $> 0.5$ MeV implied by a typical,
large X-ray burst. An uncertain fraction of these would be trapped in the corona, and the electron distribution might not extend with the same energy
dependence to 10s of MeV but it appears quite plausible that enough electrons
of the required energies are present in the flare. The minimum energy of 0.5 MeV is of course quite arbitrary; only electrons in the 10s of MeV range
and above are demanded by an ICS interpretation
of this coronal HXR source.

Close to the limb, the most favourable head-on collisions of electrons with primary photons may occur. The flux and spectrum are very close to those
given traditionally for power-law electron distributions and isotropic primary photons \citep[e.g][]{1970RvMP...42..237B,2008A&ARv..16..155K}, with modifications resulting
mostly from the presence of an upper electron cutoff energy. The number and energy distribution of electrons found
above are close to those that would be found using the traditional results; but this would not be the case for an event further from the limb.

An interpretation in terms of positrons is also possible. The spectra shown in Figure 5 would all give approximately the necessary, hard spectrum
in the several hundred keV energy range (although, as discussed above, none has precisely power-law form). For a power-law primary proton energy distribution
with energy spectral index = 2, about $10^{32}$ protons would be needed above 1 MeV. \cite{Masson:2009} found a proton flux of $2.3 \times 10^{31}$ cm$^{-3}$ above
30 MeV for this event with proton spectral index = 3, i.e. $2 \times 10^{34}$ protons above 1 MeV. Most secondary positrons presumably stop at great depths
in the atmosphere, but we would need only a few percent of them to find their way into the corona in order to account for the coronal HXR source via ICS.

\section{Conclusions and discussion}

ICS needs extreme source parameters if it is to account on its own for the bulk of flare hard X-rays \citep{1971SoPh...18..284K,1986A&A...165..235M},
particularly when `footpoint' source morphology points to an origin in the dense chromosphere. Our work does not revise this view, just points
out that ICS might be important for understanding sources in the tenuous corona. We have seen that very modest numbers of electrons or positrons at
relativistic energies could account for already observed coronal HXR sources,
even in regions so tenuous that a conventional bremsstrahlung interpretation would become problematic. Electrons would
need to be accelerated into the 100 MeV energy range; positrons are automatically produced with the necessary energies as long as there are $\sim$ GeV protons
to produce them in the first place. The electron distribution needed to account for HXR bursts, extended into the 100 MeV energy range, would include
enough relativistic electrons that only a small fraction of them would need to be found in the corona to account for at least one, observed coronal
HXR source. Moreover, electrons might be accelerated to relativistic energies via a process distinct from the main flare energy release, as appears to occur
in the Earth's magnetosphere \citep{2001JGR...10619169B}.

How might we distinguish these sources from conventional bremsstrahlung HXRs? First of all, they may be expected from locations where the ambient density seems
too low for a conventional, bremsstrahlung interpretation. As we have seen, coronal ICS sources should be brightest near the solar limb. If many sources like
those described in \cite{2008ApJ...678L..63K} can be detected, an ICS interpretation would imply a strong centre-to-limb variation. Simultaneous
observations from two, widely separated spacercraft \citep[e.g.][]{2008ApJ...678L..63K} would reveal quite different fluxes and spectra.  The
spectra will always be very hard, possibly also extending to soft X-ray and EUV ranges in a continuous way difficult to account for by other means.
Observations of co-spatial radio radiation would have very different spectral properties in the bremsstrahlung and ICS cases.

Do we need to contain electrons in the corona to produce such sources? The calculations above assume that an isotropic population of electrons
is instantaneously present in the source region. Radio observations show coronal containment
of high-energy (gyrosynchrotron emitting) electrons \citep{2001ApJ...557..880K,2002ApJ...580L.185M}. The overwhelming contribution to observed ICS,
however, comes from electrons moving
instantaneously towards the observer. Electrons could pass freely through the corona, following the field lines and emitting observable ICS
HXRs as they pass through the line of sight towards the observer. They would not need to be contained in the corona, and an isolated
coronal source might be more naturally explained in this way, as a consequence of relativistic beaming and source magnetic geometry.
Instantaneous numbers of electrons needed would be comparable to the numbers
found above. A more detailed treatment of electron and positron transport, not given here, would be needed to assess this possibility properly.

Our assumed isotropic electron distribution raises similar questions. We appealed to electron and positron scattering
by MHD turbulence to justify this assumption \citep[e.g.][]{1989ApJ...344..973M}. It still seems unclear if the coronal electron trapping revealed in
radio  is due to turbulence, magnetic field convergence and/or other physical factors. Electrons may be coronally contained but
anisotropic. The consequences of anisotropy are more easily addressed for our
highly relativistic electrons than e.g. the study of gyrosynchrotron radiation carried out by \cite{2003ApJ...587..823F}. The cone
of emission about the electron instantaneous direction of motion has width $\gamma^{-1}$, so the electron distribution function and the
loop geometry \citep[e.g. orientation north-south; any tilt to the vertical, etc. - cf.][]{1990A&A...232..544M} would  have to conspire to ensure that some electrons
travel more or less in the line of sight. Deduced numbers of electrons would again be similar, to order of magnitude, to those
found assuming isotropy but the range of viewing angles giving rise to an observable source would be narrower.

ICS coronal X-ray sources may already have been observed. Already well-studied
sources, like that in the Masuda flare \citep{1994Natur.371..495M} or some of those described by
\cite{2009A&A...502..665T}, might be reinterpreted in this way. In these smaller events, including the M class
Masuda flare, there are not $\gamma$-ray measurements to give any independent constraint on high-energy electrons or
positrons, however. In small flares, bright coronal HXR sources in implausibly tenuous regions would indicate the presence
of relativistic electrons or positrons.

Might ICS yield observable contributions in other wavelength ranges? Flare positrons, for example, would scatter
cm wavelength photons into the optical or near UV ranges. In the corona the primary photon number density would
be extremely low, making an observable flux highly unlikely, unless the relativistic electrons or positrons lay within
an optically thick microwave source. Such a situation would need a much more detailed evaluation of the primary radiation field
than we have carried out here, along the lines of \cite{1986A&A...165..235M}. Other possibilities, like an ICS contribution
to UV continuum, appear potentially interesting but would take place in the deeper atmosphere and would similarly
require a different treatment of the primary radiation field.

If definitively recognised in flares, ICS coronal HXR sources would open a new window on acceleration and transport of electrons and ions in the 0.1-1 GeV energy range.

\acknowledgements
ALM thanks S Krucker, H Hudson and the other participants in the ISSI International Team on ``Coronal Hard X-ray Sources in
Solar Flares" for helpful conversations and for providing the initial impetus for this work; and T Porter for useful conversations.
PCVM is supported by a UK STFC Dorothy Hodgkin's Scholarship. Solar physics research in Glasgow is supported by an STFC Rolling Grant.

\bibliographystyle{aa}
\bibliography{pcvm}

\end{document}